\documentstyle[12pt,epsfig]{article}

\newlength{\dinwidth}
\newlength{\dinmargin}
\begin{document}
\def\bold#1{\setbox0=\hbox{$#1$}%
     \kern-.025em\copy0\kern-\wd0
     \kern.05em\copy0\kern-\wd0
     \kern-.025em\raise.0433em\box0 }
\def\slash#1{\setbox0=\hbox{$#1$}#1\hskip-\wd0\dimen0=5pt\advance
       \dimen0 by-\ht0\advance\dimen0 by\dp0\lower0.5\dimen0\hbox
         to\wd0{\hss\sl/\/\hss}}
\def\lq{\left [}
\def\rq{\right ]}
\def\LL{{\cal L}}
\def\VV{{\cal V}}
\def\AA{{\cal A}}
\def\BB{{\cal B}}
\def\MM{{\cal M}}
\def\ovl{\overline}
\newcommand{\be}{\begin{equation}}
\newcommand{\ee}{\end{equation}}
\newcommand{\bea}{\begin{eqnarray}}
\newcommand{\eea}{\end{eqnarray}}
\newcommand{\nn}{\nonumber}
\newcommand{\dd}{\displaystyle}
\newcommand{\bra}[1]{\left\langle #1 \right|}
\newcommand{\ket}[1]{\left| #1 \right\rangle}
\newcommand{\spur}[1]{\not\! #1 \,}
\thispagestyle{empty}
\vspace*{1cm}
\rightline{BARI-TH/96-252}
\rightline{Napoli Preprint DSF-T-46/96}
\rightline{October 1996}
\rightline{hep-ph/9610297}
\vspace*{2cm}
\begin{center}
  \begin{LARGE}
  \begin{bf} 
Rare $B\to K^{(*)} \nu\ovl\nu$ Decays  at B Factories\\
  \end{bf}
  \end{LARGE}
  \vspace{8mm}
  \begin{large}
P. Colangelo$^a$, F. De Fazio$^{a,b}$, P. Santorelli$^c$, E. Scrimieri$^{a,b}$
  \end{large}
  \vspace{1cm}

\begin{it}
$^{a}$ Istituto Nazionale di Fisica Nucleare, Sezione di Bari, Italy\\
$^{b}$ Dipartimento di Fisica, Universit\'a di Bari, Italy \\
$^{c}$ Istituto Nazionale di Fisica Nucleare, Sezione di Napoli, Italy
\end{it}
\end{center}
\begin{quotation}
\vspace*{1.5cm}
\begin{center}
  \begin{bf}
  Abstract\\
  \end{bf}
\end{center}
\noindent
We compute the branching fraction of the decays 
$B \to K \nu \overline\nu$ and $B \to K^* \nu \overline\nu$ in the
Standard Model. We also comment on the experimental difficulties and
procedures to detect such modes at B factories. 

\vspace*{0.5cm}
\end{quotation}

\newpage
\baselineskip=18pt
\setcounter{page}{2}

Among the various flavour changing neutral current-induced 
$b-$quark decays \cite{review}, the transition
\be
b \to s \nu\ovl\nu \label{decay}
\ee
plays a peculiar role, both from a theoretical and an experimental point of 
view.

Within the Standard Model (SM) the
process (\ref{decay}) is governed by the effective hamiltonian
\be
{\cal H}_{eff} = {G_F \over \sqrt 2} {\alpha \over 2 \pi \sin^2(\theta_W)}
\; V_{ts} V^*_{tb} \; X(x_t) \; {\bar b} \gamma^\mu ( 1- \gamma_5) s \; 
{\bar \nu} \gamma_\mu ( 1- \gamma_5) \nu 
 \equiv  c_L^{SM} \; {\cal O}_L 
\label{hamil}
\ee
obtained from $Z^0$ penguin and box diagrams where the dominant contribution 
corresponds to a top  quark intermediate state. In (\ref{hamil})
$G_F$ is the Fermi constant, $\alpha$ the fine structure coupling constant
(at the $Z^0$ scale), $\theta_W$ the Weinberg angle and $V_{ij}$ are
Cabibbo-Kobayashi-Maskawa (CKM) matrix elements; $x_t=(m_{top}/M_W)^2$.
${\cal O}_L$ represents the left-left four-fermion operator
${\cal O}_L \equiv {\bar b} \gamma^\mu ( 1- \gamma_5) s \; 
{\bar \nu} \gamma_\mu ( 1- \gamma_5) \nu $.
The ${\cal O}(\alpha_s)$ contribution deriving from
two-loop diagrams is taken into account in the function $X$:
\be
X(x)= X_0(x) + {\alpha_s \over 4 \pi} X_1(x) \;\;,
\ee
where \cite{inami}
\be
X_0(x) = {x \over 8} \; \left [ {x + 2 \over x -1} + 
{3 x - 6 \over (x -1)^2} \ln x \right ]
\label{x0} 
\ee
and \cite{buchalla,buchalla1}
\bea
X_1(x) & = & { 4 x^3 - 5 x^2 -23 x \over 3 ( x-1)^2} -
{ x^4 + x^3 - 11 x^2 + x \over ( x-1)^3} \ln x  \nn \\
& + & { x^4 - x^3 - 4 x^2 - 8 x \over 2 ( x-1)^3} \ln^2 x +
{ x^3 - 4 x \over ( x-1)^2} L_2(1- x) + 
8 x {\partial X_0(x) \over \partial x} \ln x_\mu 
\label{x1} 
\eea
($L_2(1-x) = \dd\int_1^x dt \; \dd{\ln t \over 1 - t}$,
and $x_\mu= \dd{\mu^2 \over M_W^2}$, with $\mu= {\cal O}(m_{top})$).
Such a correction, using 
$m_{top}=175\pm 9\; GeV$ \cite{CDF} and $\alpha_s(m_b)=0.23$, is around
$3 \; \%$. 

The presence of a single operator governing the transition (\ref{decay}) 
is a welcomed property, since the theoretical uncertainty is only related, 
within SM,
to the value of one Wilson coefficient $c_L^{SM}$. In other cases, for example
in $b \to s \ell^+ \ell^-$, the effective 
hamiltonian consists of several terms coherently acting to determine branching 
ratios, invariant mass distributions, lepton charge and polarization 
asymmetries, etc., and the uncertainty of a set of coefficients appearing in 
interfering terms must be taken into account.
Moreover, possible New Physics (NP) effects contributing to (\ref{decay})
can only modify the SM value of the coefficient $c_L$, 
or introduce one  new right-right operator
\be
{\cal H}_{eff} \equiv c_L \; {\cal O}_L + c_R \; {\cal O}_R 
\label{NP}\ee
(${\cal O}_R \equiv {\bar b} \gamma^\mu ( 1+ \gamma_5) s \;
{\bar \nu} \gamma_\mu ( 1 + \gamma_5) \nu $), 
with $c_R$ only receiving contribution from phenomena beyond SM.
 
The process (\ref{decay}) is theoretically 
appealing also because of the absence 
of long-distance contributions, which are usually
related to the presence of four-quark operators in the effective 
hamiltonian and, e.g., 
heavily affect the process $b \to s \ell^+ \ell^-$ \cite{longdist}.
In this respect, the transition to neutrinos represents a clean process even 
in comparison with the $b \to s \gamma$ decay, where long-distance 
contributions are expected to be present, although small \cite{riaz}.

As for inclusive decays, the analysis of $B \to X_s \nu \ovl \nu$
in the framework of the expansion in the inverse heavy quark 
mass shows that the ${\cal O}(m_b^{-2})$
preasymptotic corrections to the partonic spectrum
are  negligible for all values of the squared 
momentum transferred to the neutrino pair, but
for a narrow region near the end-point \cite{preas}. 

Moreover, the exclusive $B \to K^{(*)}$ transitions induced by (\ref{hamil})
can be related to the semileptonic Cabibbo suppressed 
$B \to \pi (\rho) \ell \nu$ decays, on which first results are now available 
\cite{Bpi}. The idea is to set up a procedure for a determination of the ratio
of the CKM matrix elements appearing in $b\to s$ and $b\to u$ transition
$\dd|V_{ts}|/|V_{ub}|$ in a way safe of hadronic uncertainties 
\cite{ligeti1,odonnell}.  
 
From the experimental point of view, the analysis of the $B$ meson
decays induced by
(\ref{hamil}) has to be considered together with 
the study of the purely leptonic decay mode
$B^- \to \tau^- \ovl \nu_\tau$, where two neutrinos are also
produced in the final state. 
This analogy has been exploited \cite{nardi,alephconf}
to establish a bound on the inclusive 
$B \to X_s \nu \ovl \nu$ branching ratio
using the upper limit obtained by the ALEPH Collaboration at CERN 
\cite{alephconf,aleph}:
\be
\BB(B^- \to \tau^- \ovl \nu_\tau) \le 1.6 \times 10^{-3}  \;\;(at \; 90 \% \; 
CL) \;\;\; .
\ee
The resulting bound \cite{alephconf}
\be
\BB(B \to X_s \sum_i \nu_i \ovl \nu_i) \le 7.7 \times 10^{-4} \label{bound}
\ee 
must be compared to the SM prediction, which can be derived 
considering the ratio 
\be
{\BB(B \to X_s \sum_i \nu_i \ovl \nu_i) \over 
\BB(B \to X_c \ell \ovl \nu_\ell)} =
3\; {\alpha^2 \over 4 \pi^2 \sin^4(\theta_W) }
\left |{V_{ts} \over V_{cb}}\right |^2 
{X(x_t)^2 \over \eta_0 \; f(m_c/m_b)} \; \overline{\eta}
\label{brratio}
\ee
where the theoretical uncertainty related to 
the $m_b^5$ factor disappears. In eq.(\ref{brratio}) the factor $3$
accounts for the sum over the three neutrino species. Using the
phase space factor $f(m_c/m_b)\simeq 0.44$, the QCD correction factors 
$\eta_0 \simeq 0.87$ and 
$\overline{\eta} =  1 + \dd{2 \alpha_s(m_b) \over 3 \pi}\; 
\left( \dd{25 \over 4} - \pi^2 \right) \simeq 0.83$ \cite{buchalla1}, 
and the experimental measurement 
$\BB(B \to X_c \ell \ovl \nu_\ell)=(10.23 \pm0.30)\times 10^{-2}$ (at 
$\Upsilon(4S)$)
$\left[(10.95 \pm0.32)\times 10^{-2}\right]$ (at $Z^0)$ \cite{bsemil},
one gets the SM prediction for the rate of $B \to X_s \nu \ovl \nu$:
\be
\BB(B \to X_s \sum_i \nu_i \ovl \nu_i) = (4.52\pm0.17)\times 10^{-5} \;\;
\left [(4.84\pm0.14)\times 10^{-5}\right] 
\;\; \left({|V_{ts}|\over |V_{cb}|}\right)^2 \;\;\;.\label{inclusive}
\ee 

The two predictions (\ref{inclusive}) 
refer to the measurements of the semileptonic branching ratio, performed at
$\Upsilon(4S)$ and at LEP, respectively.
The comparison of (\ref{inclusive}) with the bound (\ref{bound})
gives an insight into the necessary improvement of the experimental 
facilities.
Such an improvement will be reached in the dedicated $B$-physics
experiments planned for the next future, such as CLEO III at the Cornell
$e^+ e^-$ storage ring \cite{cleo3}, and BaBar \cite{babar} and 
Belle  \cite{belle} asymmetric 
B-Factories now under construction at SLAC and KEK laboratories, respectively, 
which are expected to access $B$ meson decay modes with
branching ratios less than $10^{-5}$. Therefore, the study of 
$b\to s \nu \ovl \nu$, together with the search for
$b\to s \ell^+ \ell^-$ and $b\to s \; gluon$ processes, with a refinement of the
measurement of $B \to X_s \gamma$ 
\footnote{
The decay $b \to s \gamma$ is the only 
flavour changing neutral current $b \to s$ transition observed so far 
\cite{bsgamma};
within the errors, the experimental results, both in the 
inclusive and exclusive modes, are in agreement  
with the SM predictions, and already constrain a number of its extensions
\cite{bsgrev}.} and with the investigation of $B \to X_d \gamma$ 
will allow to exploit a complete program to test the SM properties 
at the loop level and constrain various new physics scenarios
\cite{bertolini}.

The first attempt to experimentally access
the decay (\ref{decay}) will be through the exclusive modes,  
which will be better investigated at B factories. Among such modes, the 
channels $B \to K^{(*)} \nu \ovl \nu$ are the prime candidates to 
look for, and therefore it is necessary to characterize them, by computing
the fraction of the inclusive 
branching ratio represented by such exclusive modes, and the quantities 
that possibly will be used in the experimental analyses, 
for example the distribution of missing energy in the final state.
This kind of analyses will be challenging at the future 
experimental facilities; however, it is possible
that they will be successfully carried out, mainly at the asymmetric 
B-factories.

Let us first consider the channel $B\to K\nu\ovl\nu$.
In order to compute it, we need the matrix element of the effective
hamiltonian (\ref{hamil}) between the states of the initial $B$ particle 
and the final particles $K, \nu, \ovl \nu$. 
The hadronic transition $B \rightarrow K$ mediated by the vector
${\bar s} \gamma_\mu b$ current can be parameterized in terms of 
form factors, according to the notation in \cite{bsw}
\begin{equation}
<K(p^\prime)|{\bar s} \gamma_\mu b |B(p)>=(p+p^\prime)_\mu F_1(q^2) 
+{M_B^2-M_K^2 \over q^2} q_\mu \left [ F_0(q^2)-F_1(q^2)\right ]\;,
\label{f0} 
\end{equation}
\noindent 
where $q=p-p^\prime$ is the momentum transfer to the lepton pair,
and the condition $F_1(0)=F_0(0)$ has been imposed to remove the 
singularity at $q^2=0$ in (\ref{f0}).\\
The matrix element in eq. (\ref{f0}) accounts for the
long-distance interaction of quarks and gluons in the mesons,  and
must be computed by a non-perturbative approach. 
The result of 
three-point function QCD sum rules for $F_1$
 \cite{noi} \footnote{Other determinations of the form factors have 
been obtained by light-cone sum rules \cite{absimma}, quark models 
\cite{melikov}, lattice QCD \cite{lattice} and $\chi$HQET \cite{nardulli}.}, 
in the range of squared momentum transfer 
$0 \le q^2 \le 15 \; GeV^2$ (where the method can be meaningfully applied)
can be fitted by a polar $q^2$ dependence 
\be
F_1(q^2) = {F_1(0) \over 1 - {q^2 / M_P^2} }\;, \label{f1}
\ee
with $F_1(0)= 0.25 \pm 0.03$ and $M_P= 5 \; GeV$.

The missing energy distribution in the decay $B \to K \nu \ovl \nu$
can be computed using such $F_1$ (extrapolated up to $q^2_{max}=(M_B-M_K)^2$);
defining $E_{miss}$ the energy of the neutrino pair 
in the B rest frame, and adopting the dimensionless variable
$x=E_{miss}/M_B$, we obtain, in the kinematically allowed range of $x$
\be
{1 - r \over 2} \le x \le 1 - \sqrt r
\label{limits}
\ee
($r=M_K^2/M_B^2$), the result:
\be
{d \Gamma(B \to K \nu \ovl \nu) \over dx} = 
3\;{ |c_L + c_R|^2 \,|F_1(q^2)|^2 \over 48 \pi^3 M_B} 
\sqrt{\lambda^3(q^2, M_B^2, M_K^2)}\;,
\label{scal}
\label{e:specK}
\ee
where $q^2=M_B^2 (2 x- 1)+M_K^2$. Eq.(\ref{e:specK}) shows that 
a possible NP interaction modifying the 
effective hamiltonian from (\ref{hamil}) to (\ref{NP}) only changes the 
normalization of the spectrum.
 
In fig.1 the missing energy spectrum, eq.(\ref{e:specK}), is plotted 
using (\ref{f1}) and fixing the Wilson 
coefficients $c_L$ and $c_R$ to the SM values
$c_L=\left |c_L^{SM}\right|\,=\,2.7\,\, \dd \left(|V_{ts}|/0.04 \right)
\times\,10^{-9} \; GeV^{-2}$ and $c_R=0$. The predicted branching ratio,
using $\tau(B^-)=(1.65 \pm 0.04) \times 10^{-12} \; s$, is 
\be 
\BB(B^- \to K^- \sum_i \nu_i \ovl \nu_i)\,=\;(2.4\pm 0.6)\,
\left({|V_{ts}| \over 0.04}\right)^2 \times\,10^{-6}\;,
\label {bknunu}
\ee
which corresponds to the ratio 
\be
R_{K}= {\BB(B \to K \nu \ovl \nu) \over \BB(B \to X_s \nu \ovl \nu)}\,=\, 
(5.2\pm1.3)\,\times\,10^{-2} 
\;\;\left[(4.9\pm1.2)\,\times\,10^{-2}\right].
\label{rk}
\ee

For the decay $B \to K^* \nu \ovl \nu$,
the hadronic matrix element 
can be parameterized in terms of form factors as follows:
\begin{eqnarray}
<K^*(p^\prime,\epsilon)|{\bar s} \gamma_\mu (1-\gamma_5) b |B(p)>&=&
\epsilon_{\mu \nu \alpha \beta} \epsilon^{* \nu} p^\alpha p^{\prime \beta}
{ 2 V(q^2) \over M_B + M_{K^*}}  \nonumber \\
&-& i \left [ \epsilon^*_\mu (M_B + M_{K^*}) A_1(q^2) -
(\epsilon^* \cdot q) (p+p')_\mu  {A_2(q^2) \over (M_B + M_{K^*}) } 
\right. \nonumber \\ 
&-& \left. (\epsilon^* \cdot q) {2 M_{K^*} \over q^2}   
\big(A_3(q^2) - A_0(q^2)\big) 
q_\mu \right ]\,,
\label{a1}
\end{eqnarray}
\noindent 
where $A_0(0)=A_3(0)$, and
$A_3(q^2) = \dd\frac{1}{2\,M_{K^*}}
\left[\left(M_B + M_{K^*}\right)\,A_1(q^2) + 
\left(M_{K^*}-M_B\right)\,A_2(q^2)\right]$.

The three-point QCD sum rule results
for $V, A_1$ and $A_2$ (the only relevant form factors 
for a decay into massless neutrinos) are
\cite{noi} 
\be
V(q^2) = {V(0) \over 1 - q^2 / M_P^2 }\;,
\ee
with $V(0)= 0.47 \pm 0.03$ and $M_P= 5 \; GeV$, and
\be
A_i(q^2) = A_i(0) (1 + \beta_i q^2)\;,
\ee
with $A_1(0)= 0.37 \pm 0.03$, $\beta_1=-0.023 \; GeV^{-2}$,
$A_2(0)= 0.40 \pm 0.03$, $\beta_2=0.034 \; GeV^{-2}$.

From these form factors it is easy to derive the missing energy
distribution corresponding to the longitudinally and transversely polarized 
$K^*$:
\be
{d \Gamma_L \over dx} = 3\,{ |c_L - c_R|^2 \over 24 \pi^3}
{ |\vec p^{~\prime}| \over M_{K^*}^2} 
\left [ (M_B + M_{K^*})(M_B E'-M_{K^*}^2) A_1(q^2) - {2 M_B^2 \over M_B + 
M_{K^*}} |\vec p^{~\prime}|^2 A_2(q^2) \right ]^2\,,
\label{long}
\ee
and
\be
{d \Gamma_{\pm} \over dx} = 3 { |\vec p^{~\prime}| q^2 \over 24 \pi^3}
\left | (c_L + c_R) { 2 M_B |\vec p^{~\prime}| \over M_B + M_{K^*}} V(q^2) \mp
(c_L - c_R) (M_B + M_{K^*}) A_1(q^2) \right |^2\,
\label{tran}
\ee
where $\vec p^{~\prime}$ and $E'$ are the $K^*$ three-momentum and energy in 
the $B$ meson rest frame.
The missing energy distributions are plotted in fig.2, 
using the SM values of $c_L$ and $c_R$ as
for $B\rightarrow K \nu \ovl \nu$. 
 Integrating over the full spectrum we obtain the prediction
\be
\BB(B^- \to K^{-*} \sum_i \nu_i \ovl \nu_i)\,=\,(5.1\pm0.8) 
\left({|V_{ts}| \over 0.04}\right)^2 \times10^{-6}\;,
\label {bksnunu}
\ee
which corresponds to the ratio 
\be
R_{K^*}= {\BB(B \to K^{*} \nu \ovl \nu) \over \BB(B \to X_s \nu \ovl \nu)}
\,=\,(1.2\pm 0.2)\times10^{-1} \;\; [(1.1\pm 0.2)\times10^{-1}] \;\;\;.
\label{rks}
\ee
The prediction (\ref{bksnunu}) must be compared to the upper bound
obtained by DELPHI Collaboration at LEP \cite{delphi}:
$\BB(B^0_d \to K^*(892)^0 \nu \ovl \nu)\,<\,1.0 \times 10^{-3}$
(at $90 \; \%$ CL).

The results (\ref{bknunu}) and (\ref{bksnunu}) 
are smaller than the estimates obtained in \cite{aliold} assuming a heavy 
strange quark. Moreover, (\ref{bknunu}) and (\ref{bksnunu}) 
show that the branching ratio
of $B \to K^{(*)} \sum_i \nu_i \ovl \nu_i$ is larger by a factor of
five than the analogous $b \to s$ channels  $B \to K^{(*)} \ell^+ \ell^-$
(excluding the long distance contribution from the 
conversion of $J/\psi, \psi'$ resonances).
Finally, (\ref{rk}) and (\ref{rks}) imply that the exclusive 
$B \to K \nu \ovl \nu$ and
$B \to K^* \nu \ovl \nu$ 
decays represent a fraction of 
$\simeq 20 \%$
of the inclusive rate, with some correspondence with the
$B\to K^* \gamma$ channel:
$\tilde R_{K^*}= \dd{\BB(B \to K^* \gamma) \over \BB(B \to X_s \gamma)}=0.186
\pm0.060$ \cite{cleo}. In the case of the radiative transition 
it was suggested \cite{noi1} 
that a large contribution to the inclusive width should be due to the channel 
$B \to K_1 \gamma$, $K_1$ being a $1^+$ orbital excitation of 
$K^*$. It could be interesting to investigate whether this is also the case of 
$B \to X_s \nu\ovl\nu$, searching for a $K (n\pi)$ final state in the mass 
range $1.2-1.4 \; GeV$.

The experimental search for  
$B \to K^{(*)} \nu \ovl \nu$ decays can be performed by looking for 
events with large missing energy, together with an opposite side
fully reconstructed B meson.
In the case of $K^*$ at an asymmetric B factory, 
the vertex constraint obtained from the $K^*$
decay products, together with a separation of the two B decay vertices,
would provide an efficient background rejection. 
The hermiticity features of the 
detectors and the performances of the $K^{(*)}$ identification,
together with the integrated luminosity
corresponding to the full period of data taking, will be fundamental for 
a successful detection of such interesting decays. 
It is possible that the requirement of a complete reconstruction of the 
opposite-side B meson 
can be relaxed \cite{deshpande}, and that a sufficient 
background rejection can be obtained by only imposing some  
energy and momentum conservation conditions.
To check such a procedure, a Monte Carlo generator 
using a reliable set of form factors, as the set reported here,
is required
to simulate events in the different experimental environments.

\vspace{2cm}
\noindent
{\bf Acknowledgments} \\
\noindent
We thank M.Giorgi, G.Nardulli, A.Palano and N.Paver for discussions.

\clearpage

\hskip 3 cm {\bf FIGURE CAPTIONS}
\vskip 1 cm

\noindent {\bf Fig. 1}\\
\noindent
Missing energy distribution in the decay $B \to K \nu \ovl \nu$. The sum over 
the three neutrino species is understood.

\noindent {\bf Fig. 2}\\
Missing energy distribution in the decay $B \to K^* \nu \ovl \nu$.
Notations as in fig.1.

\begin{figure}[p]
\centerline{\epsfig{file=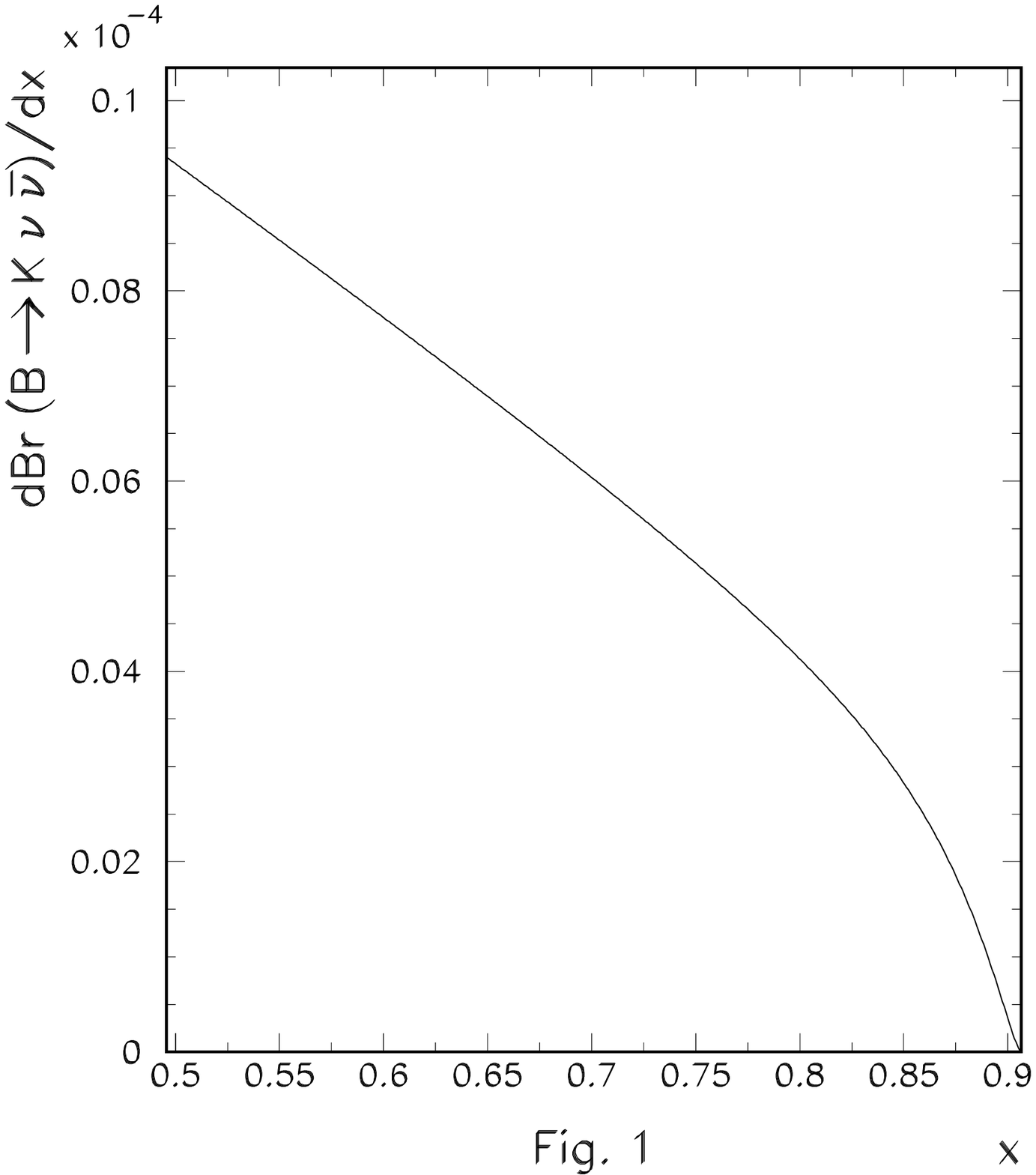,height=10cm}}
\label{fig:fig1}
\end{figure}

\vspace*{1cm}

\begin{figure}[p]
\centerline{\epsfig{file=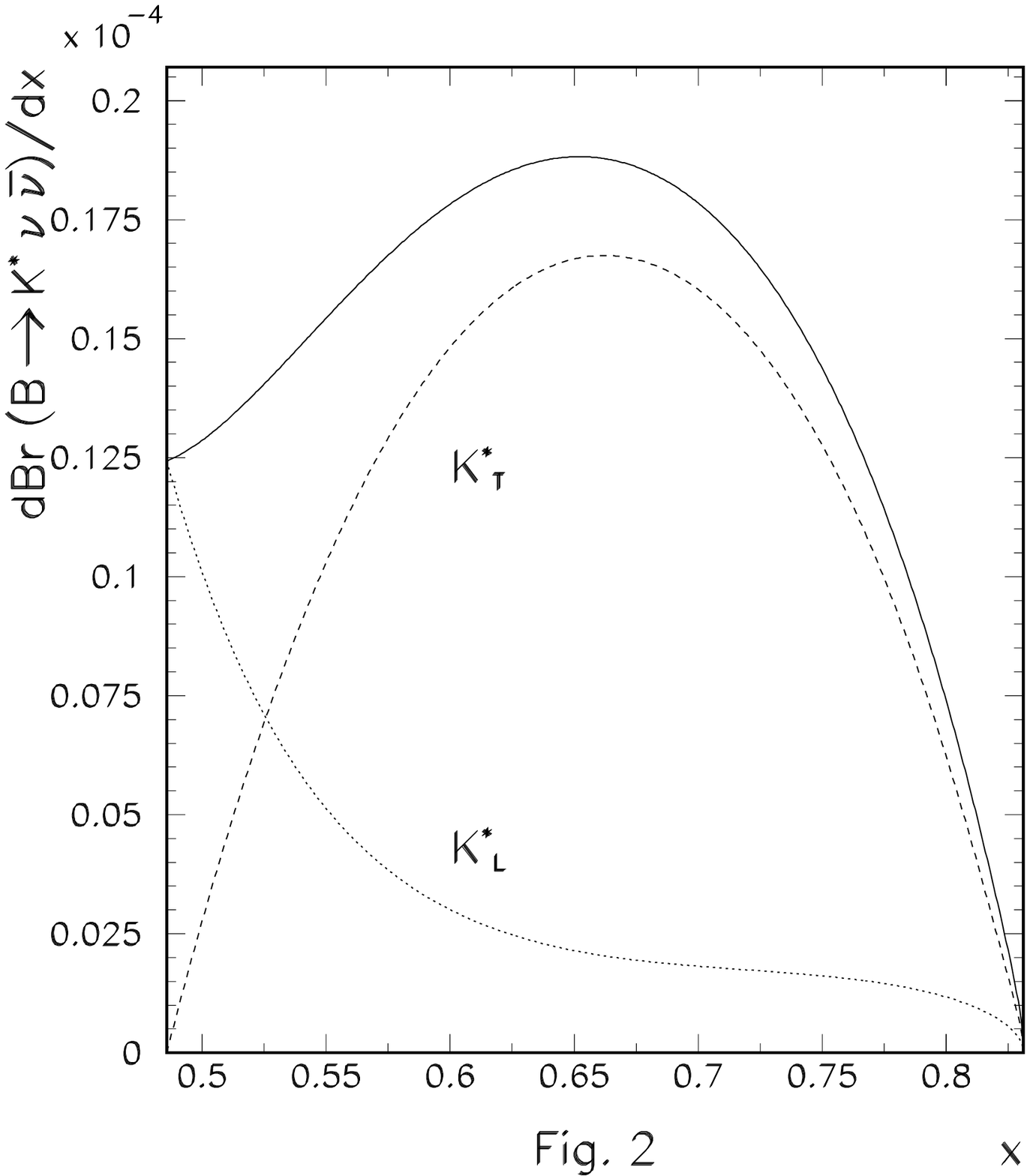,height=10cm}}
\label{fig:fig2}
\end{figure}


\begin{thebibliography}{99}

\bibitem{review} 
For a recent review see:
A.Ali, preprint DESY 96-106, to appear in the Proceedings of XX International
Nathiagali Summer College on Physics and Contemporary Needs, edited by
Riazuddin and K.A.Shoaib, (Nova Science Publishers, New York),
hep-ph/9606324.

\bibitem{inami}
T.Inami and C.S.Lim, Prog. Theor. Phys. 65 (1981) 287.

\bibitem{buchalla}
G.Buchalla and A.J.Buras, Nucl. Phys. B400 (1993) 225.

\bibitem{buchalla1}
For a review see:
G.Buchalla, A.J.Buras and M.E.Lautenbacher, report MPI-Ph/95-10, 
TUM-T31-100/95, Fermilab-Pub.95/305-T, SLAC-PUB 7009, hep-ph/9512380.

\bibitem{CDF} 
F.Abe {\it et al.}, CDF Collaboration, 
Phys. Rev. Lett. 73 (1994) 225, 
Phys. Rev. D 50 (1994) 2966, 
Phys. Rev. D 51 (1995) 4623,
Phys. Rev. Lett. 74 (1995) 2626,
Phys. Rev. D 52 (1995) R2605;
S. Abachi {\it et al.}, D0 Collaboration, 
Phys. Rev. Lett. 74 (1995) 2632.

\bibitem{longdist}
C.S.Lim, T.Morozumi and A.I.Sanda, Phys. Lett. B 218 (1989) 343; 
N.G. Deshpande, J.Trampetic and K.Panose, Phys. Rev. D 39 (1989) 1461;
P.J. O' Donnell and H.K.K. Tung, Phys. Rev. 43 (1991) R2067;
N.Paver and Riazuddin, Phys. Rev. D 45 (1992) 978.

\bibitem{riaz}
P.Colangelo, G.Nardulli, N.Paver and Riazuddin, Zeit. Phys. C - 
Particles and Fields 45 (1990) 575. 

\bibitem{preas}
A.F.Falk, M.Luke, and M.J.Savage, Phys. Rev. D 53 (1996) 2491.

\bibitem{Bpi}
CLEO Collaboration, J.P.Alexander et al., 
Phys. Rev. Lett. 77 (1996) 5000.

\bibitem{ligeti1} 
Z.Ligeti and M.B.Wise, Phys. Rev. D 53 (1996) 4937.

\bibitem{odonnell} 
P.J. O' Donnell and G.Turan, preprint UTPT-96-6, hep-ph/9604208 (June 1996).

\bibitem{nardi} 
Y.Grossman, Z.Ligeti and E.Nardi, Nucl. Phys. B 465 (1996) 369, 
hep-ph/9510378 (erratum).

\bibitem{alephconf} 
ALEPH Collaboration, Paper PA10-019 Contributed to the International Conference 
on High Energy Physics, Warsaw, Poland, 25-31 July, 1996.

\bibitem{aleph} 
D.Buskulic et al., ALEPH Coll., Phys. Lett. B 343 (1995) 444.

\bibitem{bsemil}
J.Richman, talk given at the International Conference ICHEP '96,
Warsaw 25-31 July 1996, to appear in the Proceedings.

\bibitem{cleo3} 
CLEO Collaboration, S.Kopp, talk given at the "Beauty 96" Conference, 
Roma 17-21 June 1996, to appear in the Proceedings.


\bibitem{babar} 
BaBar Collaboration, "Letter of Intent for the Study of $CP$
Violation and Heavy Flavour Physics at PEP-II", SLAC-443 (June 1994).
   
\bibitem{belle}
BELLE Collaboration, M.T.Cheng et al., KEK Preprint 94-2 (April 1994).

\bibitem{bsgamma} 
CLEO Collaboration, R.Ammar et al., Phys. Rev. Lett. 71 (1993) 674;\\
CLEO Collaboration, M.S. Alam et al., Phys. Rev. Lett. 74 (1995) 2885.

\bibitem{bsgrev}
See, for example, J.H.Hewitt, in {\it Spin Structure in High Energy Processes}, 
Proceedings of the 21st SLAC Summer Institute on Particle Physics, Stanford, 
California, edited by L. De Porcel and C.Dunwoodie (SLAC Report No. 444, 
Stanford, 1994) p.463.

\bibitem{bertolini}
S.Bertolini, F.Borzumati, A.Masiero and G.Ridolfi, Nucl. Phys. B353 (1991) 591;
\\
B.Holdom and M.V.Ramana, Phys. Lett. B 365 (1996) 309.

\bibitem{bsw}
M.Wirbel, B.Stech and M.Bauer, Zeit. Phys. C 29 (1985) 637.

\bibitem{noi}
P.Colangelo, F.De Fazio, P.Santorelli and E.Scrimieri, Phys. Rev. D 53 (1996) 
3672.

\bibitem{absimma}
A.Ali, V.M.Braun and H.Simma, Z. Phys.  C 63 (1994) 437.

\bibitem{melikov}
D. Melikov and N. Nikitin, hep-ph/9609503.

\bibitem{lattice}
APE Collaboration, C.Allton et al., Phys. Lett. B 345 (1995) 513; Phys. Lett. B 
365 (1996) 275; UKQCD Collaboration, K.Bowler et al., Nucl. Phys. B 447 (1995) 
425; Phys. Rev. D 51 (1995) 4955.

\bibitem{nardulli}
R.Casalbuoni, A.Deandrea, N.Di Bartolomeo, R.Gatto, F.Feruglio and G.Nardulli, 
Phys. Lett. B299 (1993) 139.

\bibitem{delphi}
DELPHI Collaboration, W.Adam et al., preprint CERN-PPE/96-67 (May 1996).

\bibitem{aliold}
A.Ali, C.Greub and T.Mannel, Proceedings of the ECFA Workshop on a European 
B-Meson Factory, edited by R.Aleksan and A.Ali, ECFA 93/151, DESY 93-053.
		
\bibitem{cleo}
CLEO Collaboration, R.Ammar et al., preprint CLEO CONF 96-05, 
ICHEP 96 PA05-093.

\bibitem{noi1}
P.Colangelo, C.A.Dominguez, G.Nardulli and N.Paver, Phys. Lett. B 317 
(1993) 183.

\bibitem{deshpande}
A.Schwartz, N.G.Deshpande and J.Urheim, Phys. Rev. D 44 (1991) 291;
Proceedings of the 1990 DPF Summer Study on High Energy Physics, Snowmass, 
Colorado, edited by E.L.Berger (World Scientific, Singapore, 1992) pag.278. 

\end{thebibliography}
\end{document}